# An Optimized Binning and Probabilistic Slice Sharing Technique for Motion Correction in Abdominal DW-MRI

Michelle Su[1], Cemre Ariyurek[1], Serge Vasylechko[1], Onur Afacan[1], Sila Kurugol[1]

[1]Computational Radiology Laboratory, Department of Radiology, Boston Children's Hospital, Harvard Medical School, Boston, MA, United States

**Abstract.** Diffusion-weighted magnetic resonance imaging (DW-MRI) is a powerful, non-invasive tool for detecting and characterizing abdominal lesions to facilitate early diagnosis, but respiratory motion during a scan reduces image quality and accuracy of quantitative biomarkers. Respiratory binning, which groups image slices into motion phase bins based on a respiratory navigator signal, can help mitigate motion artifacts. However, in DW-MRI, the standard binning technique often generates volumes with missing slices along the superior-inferior axis. Thus, longer scans are required to obtain complete volumes without gaps. In this study, we proposed a new binning technique to minimize missing slices without increasing scan time, aiming for more efficient motion correction. We first designed an algorithm using dynamic programming and prefix sum approaches to optimize the initial binning of the respiratory signal and MR images. Then, we developed a probabilistic refinement phase, selecting some slices to belong in two neighboring bins to further reduce missing slices. We tested our two-phase technique on free-breathing abdominal DW-MRI scans from eight subjects, including one with tumors. The proposed technique significantly reduced missing slices compared to standard binning ($p<1.0\times10^{-15}$), yielding an average reduction of 81.74±7.58%. Our technique also reduced motion artifacts, improving the conspicuity of malignant lesions. Apparent Diffusion Coefficient (ADC) maps generated from free-breathing scans corrected using the proposed technique had lower intra-subject variability compared to ADC maps from uncorrected free-breathing and shallow-breathing scans ($p<0.001$). Additionally, ADC maps from shallow-breathing scans were more consistent with corrected free-breathing maps rather than uncorrected free-breathing maps ($p<0.01$). The proposed technique corrects for motion while simultaneously reducing missing slices, allowing for shorter acquisition times compared to standard binning techniques.

## 1 | Introduction

Abdominal diffusion-weighted magnetic resonance imaging (DW-MRI) has proven to enhance evaluations of tumor progression and response to oncological treatment[1,2]. DW-MRI measures the Brownian motion of water molecules, and the acquisition of volumes with varying diffusion weighting gradients (b-values) allows for calculating apparent diffusion coefficient (ADC) maps[3]. These maps serve as important biomarkers for hepatic lesions like hepatocellular carcinoma and metastases, as well as hepatic satellite lesions, renal cell carcinoma, and splenic lesions[4–8].

However, respiratory motion can significantly degrade image quality and the accuracy of ADC maps[9], particularly in pediatric patients who often exhibit deep or irregular breathing[10]. Clinical solutions for motion correction in abdominal MRI include breath-holding and respiratory gating. Breath-holding restricts scan time to a single breath[11], limiting spatial coverage, the signal-to-noise ratio (SNR), and the number of DW-MRI averages and directions. Breath-holding is also impractical for non-cooperative patients such as young children. Respiratory gating, on the other hand, acquires images during specific phases of the respiratory signal using a navigator[12,13]. This typically occurs for 30% of the scan, so gating requires longer acquisitions and may still produce images with residual motion in cases of irregular breathing[14]. In literature, various image registration-based techniques have been explored for motion measurement and correction in DW-MRI. While 2D non-rigid registration techniques address in-plane motion, it does not correct through-plane motion in the superior-inferior direction, which is most affected by breathing[15,16]. Meanwhile, 3D non-rigid registration-based techniques[17–19] face limitations due to thick DW-MR slices and contrast between b-values. 3D slice-to-volume registration (SVR) techniques have also been employed but are typically limited to rigid motion correction in certain organs like the kidneys.[20,21].

Respiratory phase binning offers a more comprehensive approach for correcting rigid and non-rigid motion, and it has proven to be effective in T1-weighted MRI[22–24]. Recently, it has been also adapted for DW-MRI[25,26]. This method involves grouping image slices into bins based on respiratory signal from a navigator, where each bin represents one respiratory phase. Slices are averaged within each bin to improve SNR, and 3D non-rigid registration aligns and averages volumes across different bins. Binning requires shorter scan times to achieve comparable SNR with traditional respiratory gating, as it acquires data during the entire respiratory cycle.

However, in comparison to other MRI sequences, DW-MRI poses unique challenges for binning. After binning, each respiratory bin for each b-value should ideally show a complete volume containing all slices. In practice, however, due to the inability to control respiration during free-breathing acquisitions, a patient's breathing may inadvertently synchronize with scan parameters like superior-inferior slice ordering or b-values. Standard binning methods typically assign an equal number of slices to each bin[27] without accounting for these DW-MRI-specific parameters. As a result, volumes in certain bins often lack slices in certain positions along the superior-inferior axis; missing slices appear as black stripes in sagittal and coronal planes. Given that each DW-MR slice is around 4 millimeters thick, consecutive missing slices can create substantial gaps, resulting in partial or entire lesions to be missed. This problem worsens when more bins are used to capture deep breathing, as missing slices increase in frequency. Although scan times could be extended to reduce missing slices, longer scans are impractical in clinical settings.

To address these issues, we propose a new binning method that significantly reduces missing slices compared to the standard binning method, improving efficiency in terms of scan time and motion correction ability. Our two main contributions are as follows:
1) An algorithm that optimizes the initial assignment of slices to motion state bins, minimizing missing slices using dynamic programming and prefix sums.
2) An algorithm that refines initial assignments by allowing certain slices to be assigned to, further reducing the occurrence of missing slices.

Additionally, we use a Pilot Tone (PT) navigator that provides more accurate tracking of physiological motion than traditional navigators[28–32]. We develop a method to select the best PT channel to use as our respiratory navigator from the multi-channel PT data using an SNR-based criteria. For fair comparisons, we utilize the same PT navigator to track the respiratory signal for both the standard and our proposed binning methods.

Our binning method aims to reduce missing slices, increase anatomical accuracy, and shorten scan times compared to standard binning while maintaining motion-correction capabilities. This approach seeks to generate motion-robust abdominal DW-MR images with high lesion conspicuity, particularly in cases of irregular or deep breathing.

## 2 | Methods

### 2.1 Abdominal DW-MRI Experiments

We conducted liver diffusion-weighted MRI (DW-MRI) experiments on 7 subjects (4 male, 3 female, age=36.7±4.2 years) following the clinical protocol after obtaining informed consent according to an IRB-approved study. Each subject underwent two scans. For the free-breathing (FB) scan, we instructed the subject to breathe regularly, allowing for irregular and deep breathing patterns. For the shallow-breathing

(SB) scan, subjects were asked to breathe as shallowly as possible while remaining relatively still. Additionally, a free-breathing DW-MRI scan was obtained on a 22-year-old patient with metastatic disease and multiple liver lesions.

DW-MR volumes were acquired using b-values of 50, 400, 800 s/mm$^2$ with 3, 3 and 4-5 averages respectively and 6 gradient directions. All scans were performed on 3T scanners (MAGNETOM PRISMA/VIDA, Siemens Healthineers, Erlangen, Germany) using free-breathing single-shot echo planar imaging (EPI) sequence. The following image parameters were used: echo time (TE) = 78 ms, repetition time (TR) = 5200 ms, field of view = 380x310mm$^2$, in-plane resolution of 1.5x1.5mm$^2$, number of slices ($S$) = 42-52, slice thickness = 4mm, bandwidth = 2442 Hz/px, total scan time = 5.2 minutes. A Pilot Tone (PT) device[33] was placed on the patient table near the hip of each volunteer during FB scans to record motion. Raw k-space data were saved, and the corresponding navigator signal was extracted from each channel[29].

## 2.2 Pilot Tone Channel Selection

To select a PT channel most representative of physiological motion, we propose choosing the channel with the highest SNR. First, we remove any linear trend in each extracted signal to prevent the PT drift from distorting the motion signal. A linear least-squares regression line is subtracted from the data, and the data is shifted the minimum signal to 0, obtaining signal $C$. The navigator signal with the highest SNR, defined as $SNR = log_{10}(\frac{\mu_C}{\mu_{C-D}})$, is selected as the best channel. $D$ represents the denoised signal, which is the result of applying a zero-padded median filter on $C$ with a kernel size of 5, and $C-D$ represents noise.

## 2.3 Missing Slice Problem Formulation

We have a total of $N$ image slices from one scan. Each $slice_i$ possesses three fixed attributes: a corresponding value $t_i$ from the PT respiratory signal, a b-value $b_i \in \{50, 400, 800\}$, and a slice position $s_i \in \{0, 1, ..., S-1\}$ denoting the slice's location along the superior-inferior axis in one volume. Because each PT value corresponds to one image slice, when we partition the $N$ PT values into $K$ bins, we simultaneously partition $N$ slices into $K$ bins, where each bin represents one respiratory state. After binning, we obtain $3 \cdot K$ volumes, with one volume per combination of b-value and bin.

A missing slice occurs when a volume lacks an image slice for a specific slice position $s_i$. Therefore, a missing slice can be defined as the combination of $b_i$, $s_i$, and bin number where no data exists. Our goal is to partition the PT signal into $K$ bins such that each of the resulting 3x$K$ volumes contains at least one image slice for as many $s_i$ possible. Our technique minimizes the number of missing slices in two phases. The initial optimal bin assignment phase assigns each slice to one bin, creating the optimal rigid bin partitions. The probabilistic slice sharing phase allows selective slices to be shared between bins.

## 2.4 Initial Optimal Bin Assignment

Our first phase assigns each slice to one bin, designating rigid partitions in the PT signal. This phase returns the optimal assignments by finding the global minimum for the number of missing slices across $3 \cdot K$ volumes. We define our cost as the number of missing slices, so minimizing the cost is equivalent to maximizing the total number of combinations of ($b_i$, $s_i$) covered by all slices in each bin. We use a dynamic programming approach to reduce computational complexity, saving solutions to subproblems to avoid repeated computations. Our specific process is described as follows.

The minimum cost for partitioning $n$ slices into $k$ bins is denoted as $F(k, n)$. The $n$ slices are sorted by PT signal value. For each step in our dynamic programming algorithm, we focus on finding the optimal last partition point $i$ through recursive logic. We assume the solution for partitioning the first $i$ slices into $k$-1 bins is already known, allowing us to determine the solution for $n$ slices and k bins. The minimum cost for the subproblem is denoted as $F(k - 1, i)$, representing the cost for the data left of the partition $i$. The cost for the data right of the partition is denoted as $R(i, n)$, indicating the number of missing slices between $i$ and $n$. The sum of the left and right costs is the total cost.

$$F(k, n) = argmin(i)\ (F(k - 1, i) + R(i, n)) \qquad (1)$$

Our algorithm initializes two $K$x$N$ arrays to respectively store the minimum cost and the corresponding last partition $i$ for each pair of $k \in \{0, 1, \ldots, K\text{-}1\}$ and $n \in \{0, 1, \ldots, N\text{-}1\}$. We use a tabulation (bottom-up) approach to fill these arrays. In each dynamic programming step, we iterate through every possible last partition point $i \in \{0, 1, \ldots, n\text{-}1\}$ to minimize $F(k, n)$. $R(i, n)$ is queried in linear time with a $N$x3x$S$ precomputed prefix sum array *prefix*. *prefix*[$i$][ $b_i$][ $s_i$] stores the number of slices existing for each combination ($b_i$, $s_i$) in the first $i$ slices, $R(i, n)$ is the number of zeros in *prefix*[$n$][$b_i$][$s_i$]- *prefix*[$i$][ $b_i$][$s_i$]. We use a memoization (top-down) approach to reconstruct the partition points for our final $K$ and $N$. The time complexity of our algorithm is $O(K \cdot N^2)$.

## 2.5 Probabilistic Slice Sharing

Although the first phase of our technique finds optimal rigid partitions, some missing slices still exist after rigid binning. To further reduce missing slices, our second phase chooses some "shared slices" to belong in a secondary bin, where each shared slice fills an initially missing slice in the secondary bin. Such slices form an exception to initially rigid bin partitions. To start, we construct a Gaussian model for each $bin_j$, and for each combination of $slice_i$ and $bin_j$, we compute the probability density $PD(i, j)$ that $slice_i$ belongs to $bin_j$. For each $slice_i$, we compute a set of probabilities $\{p(i, 0), p(i, 1), \ldots, p(i, k\text{-}1)\}$. Each $p(i, j)$ represents the normalized probability that $slice_i$ belongs to $bin_j$, with probabilities weighted by $w_j$, the number of slices in $bin_j$. This process shares parallels with the expectation step of the expectation-maximization algorithm for Gaussian Mixture Modeling[34].

$$p_{(i,j)} = \frac{w_j \cdot PD(i, j)}{\sum_{k=1}^{K} w_k \cdot PD(i, k)} \qquad (2)$$

Recall that a missing slice is defined as a combination of $b_i$, $s_i$, and bin number where no data exists. We can thus identify candidate slices to fill each missing slice: such a slice is initially assigned to $bin_a$ but may fill a missing slice in $bin_b$. Candidate slices have the same $b_i$ and $s_i$ as the missing slice but have the initial assignment in a neighboring bin.

To select the best candidate slice, we propose a share metric to determine the potential for a slice to belong to two bins. For a candidate $slice_i$, the share metric is defined as $SM(i) = \frac{p(i,b)}{p(i,a)}$, where $bin_a$ is the initial bin and $bin_b$ is the bin with a missing slice. The candidate slice with the greatest share metric is selected as the best slice, under the constraint that a candidate cannot fill more than one missing slice. We fill a missing slice $slice_i$ with its best candidate slice (i.e. assign this slice to both $bin_a$ and $bin_b$) if $SM(i) > T$, where $T$ is the share metric threshold. In addition to filling missing slices that meet this share metric threshold, we guarantee that in a consecutive pair of missing slices, at least one of the slices in the pair was filled regardless of its share metric. This same process applies to edge missing slices ($s_0$ or $s_{S\text{-}1}$). We select the optimal $K$ as the maximum $K$ allowing for less than 2% of slices across all volumes to be missing after slice sharing.

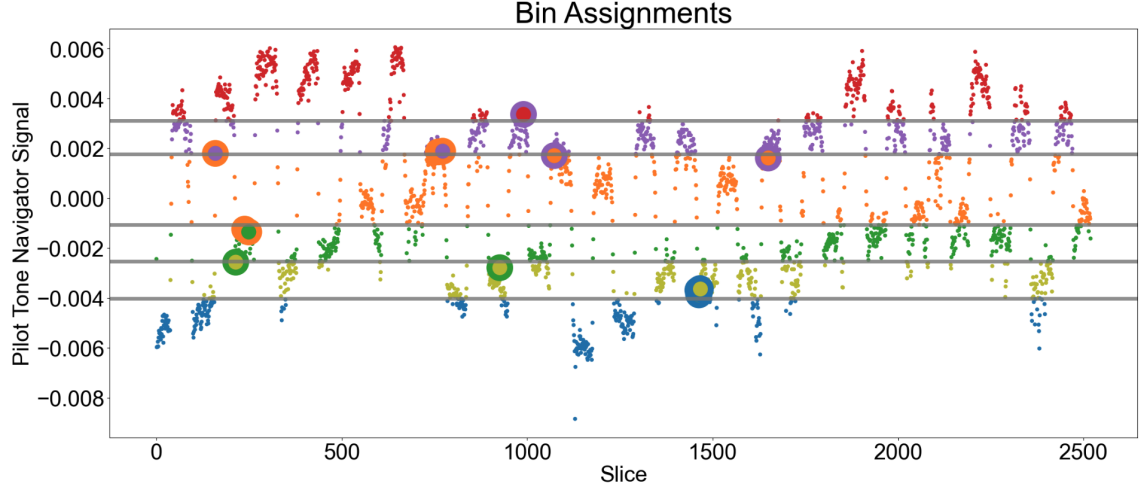

**Figure 1.** Final bin assignments using the proposed method for Subject 1 with $K = 6$ and $T = 0.1$. Gray lines denote bin partitions from the initial optimal bin assignment phase. Enlarged points represent the shared slices determined by probabilistic slice sharing, with the inner color denoting the initial bin assignment and the outer color denoting the second bin assignment.

## 2.6 Image Registration and ADC Estimation

We apply this binning process to correct for motion in free-breathing (FB) scans. We group the data based on all combinations of $b_i$, $s_i$, and bin number. We fill any missing slices remaining after the proposed binning method with linear interpolation; each slice is only interpolated from its two nearest neighbors, as we have eliminated consecutive missing slices. For combinations of $b_i$, $s_i$, and bin number that contain multiple slices, we averaged them to enhance SNR. After these processing steps, we have one slice per combination. We subsequently generate one volume for each b-value and bin, resulting in *3xK* total volumes. We perform isotropic resampling on each volume, matching the higher resolution of the axial image to the coronal and sagittal resolutions.

For each b-value, we register *K-1* floating motion state volumes to one reference motion state volume using an automated diffeomorphic non-rigid transformation[35]. We select the reference volume to be the end-expiration motion state, where the abdomen appears largest and motion is at a minimum[36,37]. This state is $bin_0$ or $bin_{K-1}$ depending on the orientation of the PT navigator signal. Following the non-rigid registration process, we resample each isotropic volume to its original dimensions. We average the *K* registered volumes for each b-value to produce a final motion-corrected free-breathing (Binning-FB) volume for each b-value.

We additionally generate uncorrected free-breathing (Uncorrected-FB) and uncorrected shallow-breathing (SB) volumes for comparison. We create uncorrected volumes by averaging raw DW-MR volumes from repetitions of the 6 diffusion directions for each b-value. For all three volume types, we use Singular Value Decomposition (SVD) to perform least squares fitting of a single exponential model decay model based on the 3 b-values at each voxel. We calculate the ADC parameter from the exponential decay model at each voxel to estimate ADC parameter maps.

## 3 | Results

### 3.1 Binning

We evaluated the number of missing slices summed across the 3x*K* binned volumes resulting from the standard and proposed binning methods for free-breathing (FB) scans. (The standard binning method assigns an equal number of slices to each bin.) In calculating subject-specific percentages of missing slices, the number of expected slices across all volumes is used as the total, which is equal to 3 (number of b-values) x $K$ (number of bins) x $S$ (number of slice positions). $K$ and $S$ vary by subject. Compared to the standard method, the proposed method yielded an average of 81.74±7.58% fewer missing slices. Without probabilistic slice sharing, the number of missing slices was reduced by an average of 36.91±9.51% compared to the standard method. A one-sided two-proportion z-test with the null hypothesis that the total proportion of missing slices across all subjects is equal using the standard and proposed method resulted in a p-value of less than $1.0\times10^{-15}$. Figure 2 shows examples of missing slices, and Table 1 shows subject-specific results for missing slices.

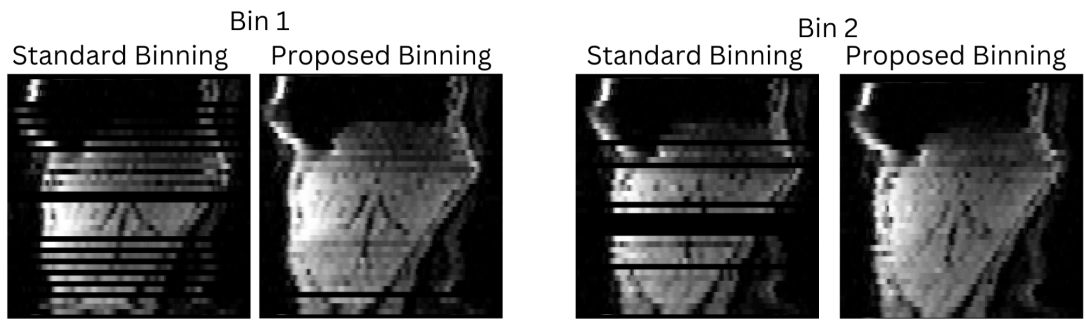

**Figure 2.** Subject 1 free-breathing (FB) b50 volumes for bins 1 and 2 in the sagittal plane generated from standard and proposed binning methods. Images obtained prior to interpolation, registration, and averaging. The proposed method eliminates isolated missing slices and large gaps with consecutive missing slices, leading to more accurate final volumes and reducing scan times.

**Table 1.** Total number and percentage of missing slices after standard binning and after each step of the proposed binning method.

| Subject | | Sub. 1 | Sub. 2 | Sub. 3 | Sub. 4 | Sub. 5 | Sub. 6 | Sub. 7 | Clinical Sub. |
|---|---|---|---|---|---|---|---|---|---|
| **Number of bins ($K$)** | | 6 | 9 | 9 | 6 | 7 | 7 | 7 | 7 |
| **Missing Slices** | Standard Binning | 44 (5.82%) | 68 (6.00%) | 67 (5.91%) | 27 (2.88%) | 74 (7.34%) | 41 (3.90%) | 65 (7.34%) | 29 (2.69%) |
| | Proposed binning: phase 1 | 19 (2.51%) | 48 (4.23%) | 42 (3.70%) | 19 (2.03%) | 42 (4.20%) | 28 (2.67%) | 46 (5.22%) | 18 (1.67%) |
| | Proposed binning: phase 2 | 4 (0.53%) | 10 (0.88%) | 7 (0.62%) | 7 (0.75%) | 13 (1.29%) | 9 (0.86%) | 10 (1.14%) | 9 (0.83%) |
| **Reduction in missing slices with proposed binning (%)** | | 90.90 | 85.29 | 89.56 | 74.07 | 82.43 | 78.05 | 80.56 | 69.0 |

### 3.3 Qualitative Evaluation of Lesion Conspicuity and Image Quality

We observed image quality for the axial and sagittal plane images of FB scans (Figures 3-5). The standard binning technique cannot produce final volumes because non-rigid registration does not handle the existence of consecutive missing slices, which are shown in Figure 2. Therefore, we compared images produced from the proposed binning technique (Binning-FB) and uncorrected images (Uncorrected-FB). Compared to Uncorrected-FB images, Binning-FB axial images (corrected with the proposed technique) consistently reduced motion artifacts surrounding the kidneys, liver, and spleen. Binning-FB axial images of the kidney and liver also showed improved conspicuity of blood vessels, and sagittal images showed reduced blurring and increased conspicuity of organ boundaries (Figure 5). The black sections near the bottom of Binning-FB sagittal images were a result of non-rigid registration transformations rather than missing slices. We additionally observed malignant lesion conspicuity in the clinical subject (Figure 3) and benign lesion conspicuity in two healthy subjects (Figure 4). For both Binning-FB and Uncorrected-FB images, the slice that provided the clearest view of the lesions was selected. In both independent b-value volumes and ADC maps, lesions that were nearly invisible in Uncorrected-FB images appeared clearly in Binning-FB images.

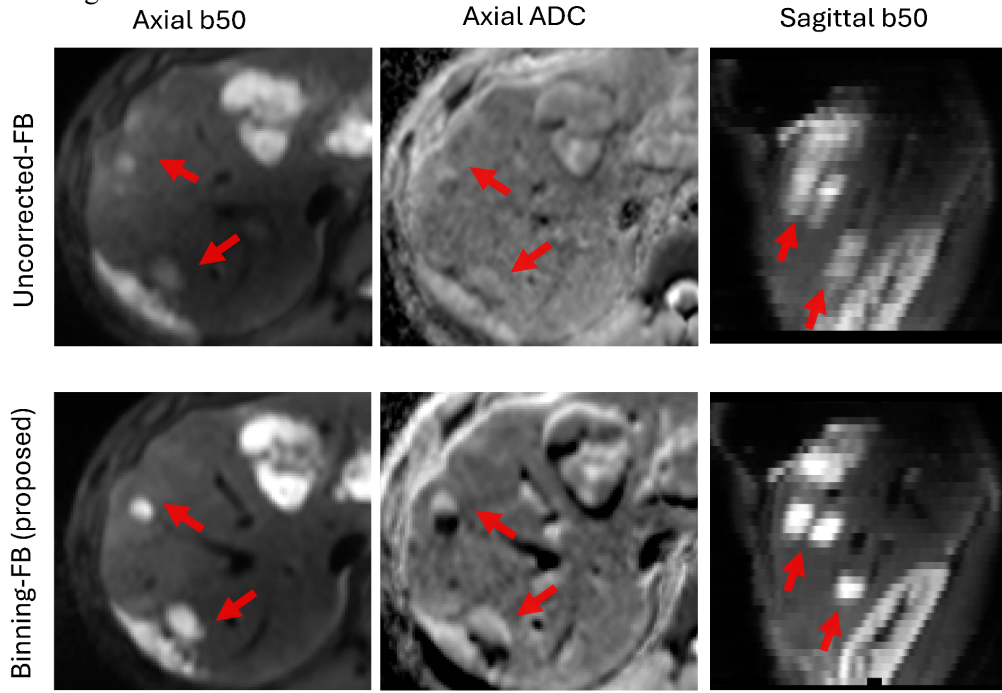

**Figure 3.** Malignant lesions in free-breathing (FB) b50 volumes and ADC maps of the clinical subject (with and without motion correction by the proposed binning method). Images obtained after registering and averaging all bins. Images corrected with the proposed method show improved lesion conspicuity, allowing for more accurate lesion characterization and treatment administration.

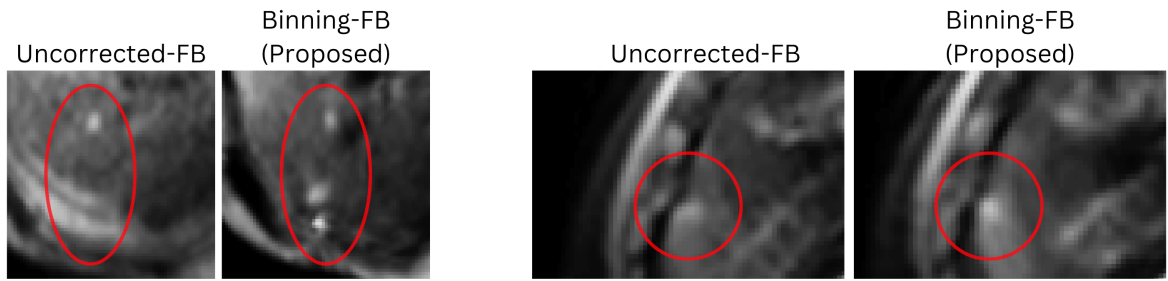

**Figure 4.** Axial view of benign lesions in Subject 6 (left) and Subject 7 (right) in free-breathing (FB) b50 volumes (with and without motion correction by the proposed binning method). Images obtained after registering and averaging all bins. Images corrected with the proposed method reveal lesions that are invisible in uncorrected images.

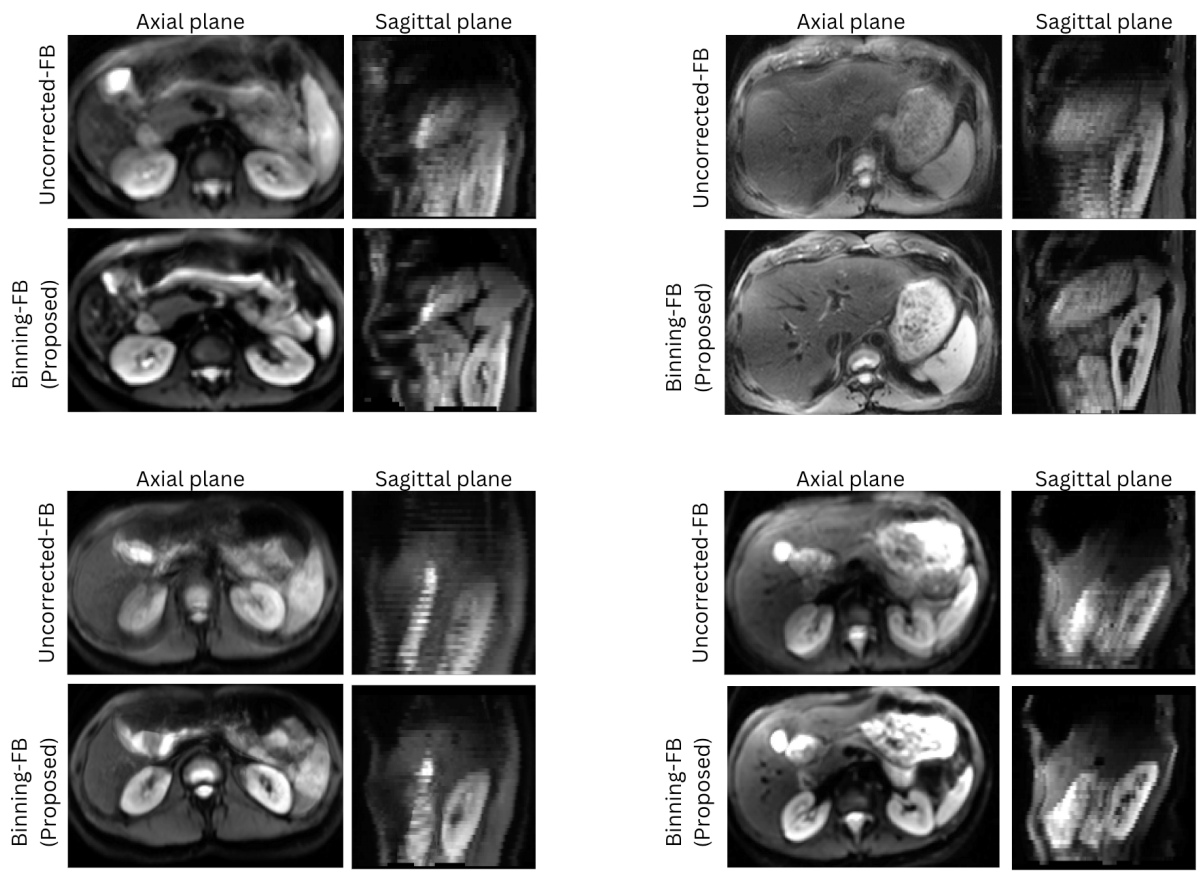

**Figure 5.** Full free-breathing (FB) b50 volumes from four non-clinical subjects in the axial and sagittal planes (with and without motion correction by the proposed binning method). Binning-FB images show increased conspicuity, reduced motion artifacts, and reduced blurring, which may allow more precise diagnoses.

### 3.2 Quantitative Evaluation of ADC Estimation

We evaluated our proposed binning method's capability for estimating ADC maps by comparing Binning-FB and Uncorrected-FB ADC values. ADC maps generated from uncorrected shallow-breathing scans (SB) were used as an additional point of comparison. We segmented three regions of interests (ROIs) in homogeneous and normal areas of the spleen, liver, and left kidney for each non-clinical subject. For these ROI segmentations, we created one 2D 64-voxel mask for the Binning-FB and Uncorrected-FB scans, and we created corresponding 64-voxel mask for the SB scans by matching biomarkers around each ROI between FB and SB scans. The violin plots in Figure 6 compare ADC values in the ROIs of each subject, indicating that Binning-FB ADC distributions show the least variation in homogeneous ROIs.

We computed two metrics based on ROI ADC values to measure variability within subject ROIs and variability across subject ROIs. To measure intra-subject variability (left half of Table 2), we calculated the coefficient of variation (CoV) of ADC within a subject's ROI. We then compared the mean of these CoVs between the three image types. The mean CoV of the 21 ROIs in Binning-FB images was 72.78% lower compared to Uncorrected-FB images and 58.51% lower compared to SB images. The mean CoV of each ROI type (spleen, liver, and kidney) is also lowest in Binning-FB images. One-sided two-sample t-tests (unequal variances) for a negative difference in mean CoV between (i) Binning-FB and SB and (ii) Binning-FB and Uncorrected-FB both yielded p-values of less than 0.001. This indicates that Binning-FB ADC values show less intra-subject variability, indicating higher precision in ADC values.

Additionally, we used a separate metric to measure inter-subject variability (right half of Table 2). We first computed the mean ADC value in a subject's ROI, and we then found the CoV of the mean ADC values across different subjects. For ROIs of each type and for all 21 ROIs combined, the CoV of mean ADC values across the 7 subjects is lower for Binning-FB than for Uncorrected-FB and SB, indicating less variability in ADC values across subjects. For the 21 ROIs, the CoV of mean ADC values across the 7 subjects in Binning-FB images is 63.91% lower compared to Uncorrected-FB images and 45.07% lower compared to SB images.

We also computed two metrics to characterize the resemblance between (i) Binning-FB images and (ii) Uncorrected-FB and SB images (Table 3). For each ROI, we computed the Wasserstein distance and the root-mean-square error (RMSE) for each of the mentioned image pairings. Mean Wasserstein distances and RMSE values were consistently lower between Binning-FB and SB compared to Uncorrected-FB and SB. On average, Wasserstein distances were 50.00% lower between Binning-FB and SB, and RMSE values were 42.68% lower between Binning-FB and SB. Four one-sided two-sample t-tests (unequal variances) for a negative difference in these metrics between (i) Binning-FB and SB images and (ii) Uncorrected-FB and SB images each yielded p-values of less than 0.01. This demonstrates that SB images tend to be more similar to Binning-FB images than Uncorrected-FB images.

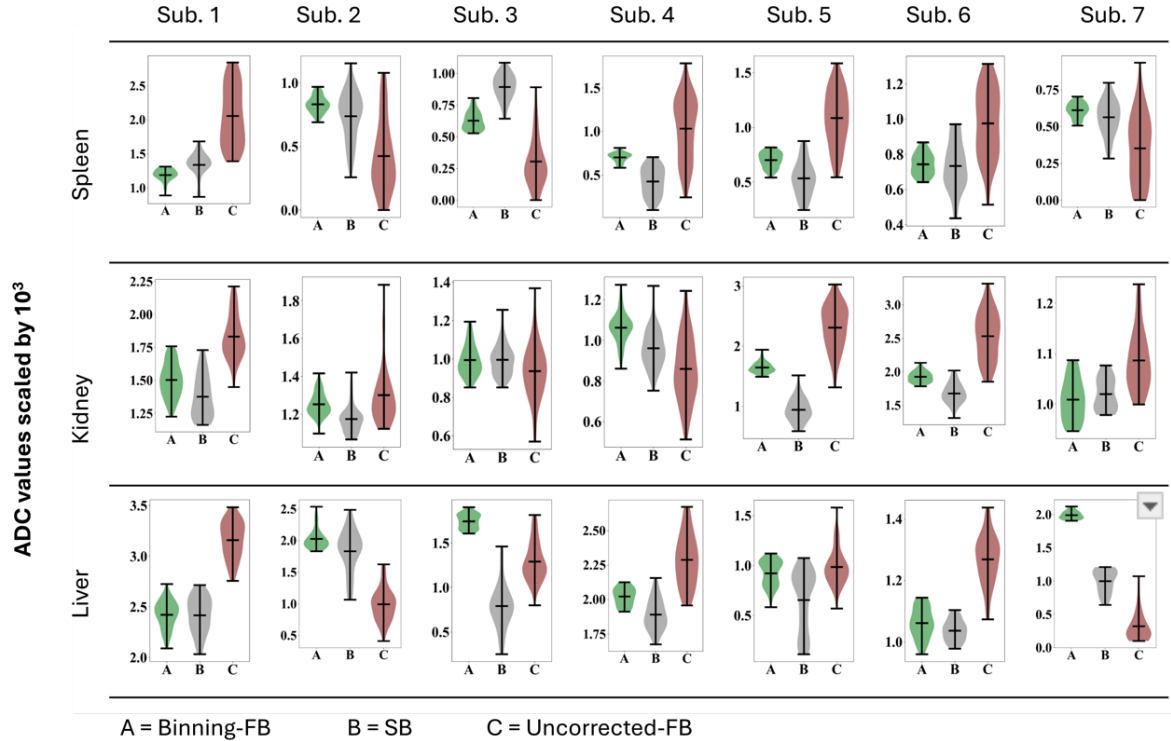

**Figure 6.** Violin plots (mean ± standard deviation) for ADC values for each of the three homogeneous ROI segmentations (spleen, liver, and left kidney) for non-clinical subjects. ADC values are compared for uncorrected shallow-breathing (SB), motion-corrected free-breathing (Binning-FB), and uncorrected free-breathing images (Uncorrected-FB). Mean ADC values are closer between SB and Binning-FB than between SB and Uncorrected-FB, elucidating the ability of the proposed motion-correction technique to render free-breathing more similar to shallow-breathing images. Standard deviation of ADC values is lowest in Binning-FB images, indicating less variability within the homogeneous ROIs.

**Table 2.** Intra-subject variability and inter-subject variability, for homogeneous spleen, liver, and kidney ROIs. Lower intra-subject and inter-subject variability is expected for more motion-robust images.

|  | Intra-subject Variability: Mean CoV of ADC in ROIs (%) | | | Inter-subject Variability: CoV of mean ADC across subjects (%) | | |
|---|---|---|---|---|---|---|
|  | SB | Binning-FB (Proposed) | Uncorrected-FB | SB | Binning-FB (Proposed) | Uncorrected-FB |
| **Spleen** | 22.43 | 8.80 | 42.65 | 40.68 | 25.68 | 69.17 |
| **Liver** | 12.77 | 7.43 | 12.25 | 21.28 | 17.86 | 27.99 |
| **Kidney** | 15.88 | 4.96 | 22.95 | 40.20 | 12.58 | 54.14 |
| **All ROIs** | 17.03 | 7.06 | 25.95 | 34.05 | 18.71 | 50.43 |

**Table 3.** Mean Wasserstein distances and root mean square error (RMSE) values between (i) Binning-FB images and SB images and between (ii) Uncorrected-FB and SB images.

|  | Mean ADC Wasserstein distance | | Mean ADC root mean square error (RMSE) | |
|---|---|---|---|---|
|  | Binning-FB (Proposed) and SB | Uncorrected-FB and SB | Binning-FB (Proposed) and SB | Uncorrected-FB and SB |
| **Spleen** | $1.59\times10^{-4}$ | $4.63\times10^{-4}$ | $1.59\times10^{-4}$ | $4.63\times10^{-4}$ |
| **Liver** | $8.92\times10^{-5}$ | $1.98\times10^{-4}$ | $8.92\times10^{-5}$ | $1.98\times10^{-4}$ |
| **Kidney** | $4.66\times10^{-4}$ | $7.68\times10^{-4}$ | $4.66\times10^{-4}$ | $7.68\times10^{-4}$ |
| **All ROIs** | $2.38\times10^{-4}$ | $4.76\times10^{-4}$ | $2.38\times10^{-4}$ | $4.76\times10^{-4}$ |

## 4 | Discussion

The proposed binning method is advantageous over the standard binning method in maintaining anatomical accuracy. The proposed method results in an average of only 0.87% total missing slices, compared to an average of 5.60% total slices missing with the standard method. Additionally, the proposed method eliminates any consecutive missing slices, thus eliminating gaps greater than 4mm thick. With standard binning, lesions within gaps are irrecoverable by interpolation; when bins are registered and averaged, even one inaccurate bin can blur a lesion in the averaged image. The proposed method addresses this issue by reducing missing slices in all bins, providing more accurate DW-MRI images for radiologist evaluation.

By reducing missing slices, the proposed method also reduces DW-MRI scan times. To reduce missing slices while still using the standard binning method, scan times would be extended unpredictably, as redundant images in combinations of $b_i$, $s_i$, and bin number that already contain data would also be acquired. The proposed binning method maximizes data utilization, making it a more viable option for clinical translation.

The proposed method is also effective in motion correction, reducing image artifacts resulting from a patient's breathing. Due to the separate issue of anatomical inaccuracies with standard binning, we focus on evaluating the proposed binning method for motion corrected. In axial images, the proposed method reduces motion artifacts around abdominal organs and increases conspicuity of lesions. Therefore, the proposed method may help clinicians better analyze lesions, enhancing diagnosis of cancer. In sagittal images, Uncorrected-FB sagittal show extensive motion in the superior-inferior direction, but such motion is eliminated or significantly reduced in Binning-FB images. This demonstrates that the DW-MRI slices are correctly ordered after motion correction.

Quantitative analysis of ADC values supports this motion correction capability. Compared to both uncorrected shallow-breathing (SB) and uncorrected free-breathing (Uncorrected-FB) images, motion-corrected free-breathing (Binning-FB) images show lower intra-subject and inter-subject variability. Since we select ROIs from homogeneous and healthy areas, we expect motion-robust images to show low variability within and across the 7 non-clinical subjects. Our results indicate that our method leads to increased stability of the ADC parameter, which is essential for helping clinicians track areas with lower ADC values to correctly detect lesions. The better quality of Binning-FB compared to SB indicates that SB scans may still be affected by motion, but the proposed binning is even more effective than attempts at shallow breathing. In other words, Binning-FB may be closer to the ground truth than SB. Although SB cannot be used as a ground truth, analysis using Wasserstein distances and RMSE calculations indicate that Binning-FB and SB show more similar ADC values than Uncorrected-FB and SB.

A limitation of the proposed method is its slower speed in comparison to the standard method. For $K = 10$ bins, the proposed method takes approximately 20 minutes, with variations depending on the length of the acquisition and the number of slices in the field of view, while the standard method takes less than one minute. However, the dynamic programming structure of the proposed method creates cumulative results for all possible numbers of $K$ from 1 to the user-specified maximum $K$. This system decreases the time spent for running different values of $K$. Additionally, this allows an automated algorithm to select the optimal number of bins $K$. With the standard method, binning must be executed for each individual $K$, and results are not cumulative.

The speed of the proposed binning method has the potential to be reduced with future work involving parallel computations. Future work can also involve optimizing the number of bins based on a subject's depth of breathing. We used higher values of $K$ in this work, but fewer bins may be used for clinical acquisitions with shorter scan times. We also intend to evaluate our method in more subjects with malignant lesions.

## 5 | Conclusion

In this study, we propose a new respiratory phase binning technique to minimize the number of missing slices in reconstructed DW-MRI volumes while simultaneously correcting for motion artifacts. Our proposed binning technique optimizes initial bin partitions and utilizes a probabilistic approach to designate slices to be shared between neighboring bins. Our technique results in significantly fewer missing slices compared to the standard uniform binning technique. This increases anatomical accuracy and allows for shorter DW-MRI acquisitions. Our technique also successfully eliminates motion artifacts and improves malignant and benign lesion conspicuity, potentially helping clinicians improve their ability to diagnose abdominal cancers.

## Acknowledgements

This work was supported partially by the National Institute of Diabetic and Digestive and Kidney Diseases (NIDDK), National Institute of Biomedical Imaging and Bioengineering (NIBIB), National Institute of Neurological Disorders and Stroke (NINDS) and National Library of Medicine (NLM) of the National Institutes of Health under award numbers R01DK125561, R21DK123569, R21EB029627, R01NS121657, R01LM013608, R01NS133228, and by the grant number 2019056 from the United States-Israel Binational Science Foundation (BSF), and a pilot grant from National Multiple Sclerosis Society under Award Number PP-1905-34002. We thank the Center for Advanced Imaging Innovation and Research (CAI2R) at NYU for supplying the 3T Pilot Tone device used in this work.

## References

1. Serai SD. Basics of magnetic resonance imaging and quantitative parameters T1, T2, T2*, T1rho and diffusion-weighted imaging. *Pediatr Radiol*. 2022;52(2):217-227. doi:10.1007/s00247-021-05042-7

2. Caro-Domínguez P, Gupta AA, Chavhan GB. Can diffusion-weighted imaging distinguish between benign and malignant pediatric liver tumors? *Pediatr Radiol*. 2018;48(1):85-93. doi:10.1007/s00247-017-3984-9

3. Charles-Edwards EM, deSouza NM. Diffusion-weighted magnetic resonance imaging and its application to cancer. *Cancer Imaging*. 2006;6(1):135-143. doi:10.1102/1470-7330.2006.0021

4. Bonekamp S, Corona-Villalobos CP, Kamel IR. Oncologic applications of diffusion-weighted MRI in the body. *J Magn Reson Imaging*. 2012;35(2):257-279. doi:10.1002/jmri.22786

5. Abugamra S, Yassin A, Abdel-Rehim ASM, Sheha DS. Apparent diffusion coefficient for differentiating between benign and malignant hepatic focal lesions. *Egyptian Liver Journal*. 2020;10(1):1-8. doi:10.1186/s43066-020-00068-2

6. Afaq A, Andreou A, Koh DM. Diffusion-weighted magnetic resonance imaging for tumour response assessment: why, when and how? *Cancer Imaging*. 2010;10 Spec no A(1A):S179-S188. doi:10.1102/1470-7330.2010.9032

7. Galea N, Cantisani V, Taouli B. Liver lesion detection and characterization: role of diffusion-weighted imaging. *J Magn Reson Imaging*. 2013;37(6):1260-1276. doi:10.1002/jmri.23947

8. Sharma K, Agarwala S, Kandasamy D, et al. Role of Diffusion Weighted MRI (DW-MR) in Detection of Satellite Lesions Not Detected with Multiphase CT Scans in Hepatoblastoma and Its Implications for Management. *Indian J Pediatr*. 2022;89(10):968-974. doi:10.1007/s12098-021-04016-9

9. Axel L, Summers RM, Kressel HY, Charles C. Respiratory effects in two-dimensional Fourier transform MR imaging. *Radiology*. 1986;160(3):795-801. doi:10.1148/radiology.160.3.3737920

10. Chavhan GB, Babyn PS, Vasanawala SS. Abdominal MR imaging in children: motion compensation, sequence optimization, and protocol organization. *Radiographics*. 2013;33(3):703-719. doi:10.1148/rg.333125027

11. Paling MR, Brookeman JR. Respiration artifacts in MR imaging: reduction by breath holding. *J Comput Assist Tomogr*. 1986;10(6):1080-1082. doi:10.1097/00004728-198611000-00046

12. Li Q, Wu X, Qiu L, Zhang P, Zhang M, Yan F. Diffusion-weighted MRI in the assessment of split renal function: comparison of navigator-triggered prospective acquisition correction and breath-hold acquisition. *AJR Am J Roentgenol*. 2013;200(1):113-119. doi:10.2214/AJR.11.8052

13. Lewis CE, Prato FS, Drost DJ, Nicholson RL. Comparison of respiratory triggering and gating techniques for the removal of respiratory artifacts in MR imaging. *Radiology*. 1986;160(3):803-810. doi:10.1148/radiology.160.3.3737921

14. Kataoka M, Kido A, Yamamoto A, et al. Diffusion tensor imaging of kidneys with respiratory triggering: optimization of parameters to demonstrate anisotropic structures on fraction anisotropy maps. *J Magn Reson Imaging*. 2009;29(3):736-744. doi:10.1002/jmri.21669

15. Glutig K, Krüger PC, Oberreuther T, et al. Preliminary results of abdominal simultaneous multi-slice accelerated diffusion-weighted imaging with motion-correction in patients with cystic fibrosis and impaired compliance. *Abdom Radiol (NY)*. 2022;47(8):2783-2794. doi:10.1007/s00261-022-03549-7

16. Son JS, Park HS, Park S, et al. Motion-Corrected versus Conventional Diffusion-Weighted Magnetic Resonance Imaging of the Liver Using Non-Rigid Registration. *Diagnostics (Basel)*. 2023;13(6). doi:10.3390/diagnostics13061008

17. Kurugol S, Freiman M, Afacan O, et al. Motion-robust parameter estimation in abdominal diffusion-weighted MRI by simultaneous image registration and model estimation. *Med Image Anal*. 2017;39:124-132. doi:10.1016/j.media.2017.04.006

18. Kurugol S, Freiman M, Afacan O, et al. Motion Compensated Abdominal Diffusion Weighted MRI by Simultaneous Image Registration and Model Estimation (SIR-ME). *Med Image Comput Comput Assist Interv*. 2015;9351:501-509. doi:10.1007/978-3-319-24574-4_60

19. Mazaheri Y, Do RKG, Shukla-Dave A, Deasy JO, Lu Y, Akin O. Motion correction of multi-b-value diffusion-weighted imaging in the liver. *Acad Radiol*. 2012;19(12):1573-1580. doi:10.1016/j.acra.2012.07.005

20. Coll-Font J, Afacan O, Hoge S, et al. Retrospective Distortion and Motion Correction for Free-Breathing DW-MRI of the Kidneys Using Dual-Echo EPI and Slice-to-Volume Registration. *J Magn Reson Imaging*. 2021;53(5):1432-1443. doi:10.1002/jmri.27473

21. Kurugol S, Marami B, Afacan O, Warfield SK, Gholipour A. Motion-Robust Spatially Constrained Parameter Estimation in Renal Diffusion-Weighted MRI by 3D Motion Tracking and Correction of Sequential Slices. *Mol Imaging Reconstr Anal Mov Body Organs Stroke Imaging Treat (2017)*. 2017;10555:75-85. doi:10.1007/978-3-319-67564-0_8

22. Shahzadi I, Siddiqui MF, Aslam I, Omer H. Respiratory motion compensation using data binning in dynamic contrast enhanced golden-angle radial MRI. *Magn Reson Imaging*. 2020;70:115-125. doi:10.1016/j.mri.2020.03.011

23. Su M, Ariyurek C, Chow J, Afacan O, Kurugol S. Enhancing Motion Correction in T1-Weighted Abdominal MRI Through Outlier Removal and Overlapping Binning. In: *Proc. Intl. Soc. Mag. Reson. Med. 32 (2024)*. ; 2024.

24. Feng L, Axel L, Chandarana H, Block KT, Sodickson DK, Otazo R. XD-GRASP: Golden-angle radial MRI with reconstruction of extra motion-state dimensions using compressed sensing. *Magn Reson Med*. 2016;75(2):775-788. doi:10.1002/mrm.25665

25. Bush M, Vahle T, Krishnamurthy U, et al. Motion Correction of Abdominal Diffusion-Weighted MRI Using Internal Motion Vectors. In: *Proc. Intl. Soc. Mag. Reson. Med. 29 (2021)*. ; 2021.

26. Vasylechko S, Ariyurek C, Afacan O, Kurugol S. Respiratory binning with PilotTone Navigator For Motion Compensated Liver DW-MRI. In: *Proc. Intl. Soc. Mag. Reson. Med. 31 (2023)*. ; 2023.

27. Jhooti P, Gatehouse PD, Keegan J, Bunce NH, Taylor AM, Firmin DN. Phase ordering with automatic window selection (PAWS): a novel motion-resistant technique for 3D coronary imaging. *Magn Reson Med*. 2000;43(3):470-480. doi:10.1002/(sici)1522-2594(200003)43:3<470::aid-mrm20>3.0.co;2-u

28. Ariyurek C, Tess W, Kober T, Kurugol S, Afacan O. Comparison of k-space center, FID and Pilot Tone navigators in abdominal motion tracking for optimal XD-GRASP reconstruction. In: *Proc. Intl. Soc. Mag. Reson. Med. 31 (2023)*. ; 2023.

29. Ariyurek C, Vasylechko S, Zhong X, et al. Pilot Tone-navigated motion estimation for liver DW-MRI. In: *Proc. Intl. Soc. Mag. Reson. Med. 31 (2023)*. ; 2023.

30. Ludwig J, Speier P, Seifert F, Schaeffter T, Kolbitsch C. Pilot tone-based motion correction for prospective respiratory compensated cardiac cine MRI. *Magn Reson Med*. 2021;85(5):2403-2416. doi:10.1002/mrm.28580

31. Solomon E, Rigie DS, Vahle T, et al. Free-breathing radial imaging using a pilot-tone radiofrequency transmitter for detection of respiratory motion. *Magn Reson Med*. 2021;85(5):2672-2685. doi:10.1002/mrm.28616

32. Falcão MBL, Di Sopra L, Ma L, et al. Pilot tone navigation for respiratory and cardiac motion-resolved free-running 5D flow MRI. *Magn Reson Med*. 2022;87(2):718-732. doi:10.1002/mrm.29023

33. Schroeder L, Wetzl J, Maier A, et al. A Novel Method for Contact-Free Cardiac Synchronization Using the Pilot Tone Navigator. In: *Proc. Intl. Soc. Mag. Reson. Med. 31 (2016)*. Accessed August 22, 2024. https://archive.ismrm.org/2016/0410.html

34. Reynolds D. Gaussian Mixture Models. *Encyclopedia of biometrics*. Published online October 18, 2018:659-663. doi:10.1007/978-0-387-73003-5_196

35. Commowick O, Wiest-Daesslé N, Prima S. Automated diffeomorphic registration of anatomical structures with rigid parts: application to dynamic cervical MRI. *Med Image Comput Comput Assist Interv*. 2012;15(Pt 2):163-170. doi:10.1007/978-3-642-33418-4_21

36. Vu KN, Haldipur AG, Roh ATH, Lindholm P, Loening AM. Comparison of End-Expiration Versus End-Inspiration Breath-Holds With Respect to Respiratory Motion Artifacts on T1-Weighted Abdominal MRI. *AJR Am J Roentgenol*. 2019;212(5):1024-1029. doi:10.2214/AJR.18.20239

37. Plathow C, Ley S, Zaporozhan J, et al. Assessment of reproducibility and stability of different breath-hold maneuvres by dynamic MRI: comparison between healthy adults and patients with pulmonary hypertension. *Eur Radiol*. 2006;16(1):173-179. doi:10.1007/s00330-005-2795-9